\documentclass[onecolumn,showpacs,showkywords,preprintnumbers,amsmath,amsfonts,mathrsfs,amssymb,preprint]{revtex4}

\usepackage[latin1]{inputenc}
\usepackage[pdftex]{graphicx}
\usepackage{epsfig}
\usepackage{graphicx}
\usepackage{dcolumn}
\usepackage{bm}
\usepackage{color}
\usepackage{natbib}
\newcommand {\be}{\begin{equation}}
\newcommand {\ee}{\end{equation}}
\newcommand {\ba}{\begin{eqnarray}}
\newcommand {\ea}{\end{eqnarray}} 

\usepackage{amsmath}  
\usepackage{amsfonts} 
\usepackage{graphicx} 
\usepackage[toc,page]{appendix}

\begin{document}

\title{A forced thermal ratchet in a memory heat bath} 

\author{O. Contreras-Vergara}
\author{N. S\'anchez-Salas}
\affiliation{Departamento de F\'isica,
Escuela Superior de F\'isica y Matem\'aticas, Instituto Polit\'ecnico Nacional, 
 Edif. 9 UP Zacatenco, CP 07738, CDMX, M\'exico.}
\author{I. P\'erez Castillo and J. I. Jim\'enez-Aquino}
\affiliation{Departamento de F\'{\i}sica, Universidad Aut\'onoma Metropolitana--Iztapalapa, 
C.P. 09340, CDMX, M\'exico.}

\date{\today}

\begin{abstract}
The present work studies a non-Markovian forced thermal ratchet model on an asymmetric periodic potential. The Brownian dynamics is described by a generalized Langevin equation with an Ornstein-Uhlenbeck-type friction memory kernel. We show that for the case of a time-dependent driving force, also in the form of an Ornstein-Uhlenbeck-like process, an exact expression of the probability current can be  derived. We also obtain the behavior of the particle's average rate of flow as a function of the external amplitude force and of the bath temperature when the driving force behaves as a square wave modulation. All our results are compared with those obtained in the Markovian case and we find, fairly remarkably, that in some cases a friction memory kernel results in an enhancement of the current.
\end{abstract}
\maketitle 

\section{Introduction} 
Thermal ratchets are  devices in which thermal noise acting on a Brownian particle is rectified and harnessed to do useful work. In this sense, a thermal ratchet can be considered as a directed transport phenomenon where the thermal noise plays a fundamental role. Even though  directed transport in a system with a single thermal bath is forbidden by the Second Law of Thermodynamics, it can be achieved under certain conditions as, for instance, when the system is coupled to two different heat baths, when it is  periodically driven, by adding non-linearities, by incorporating stochastic temporal or deterministic driving forces without preferential direction, by considering asymmetric and periodic potentials, etc. Feynman \cite{Feynman1966} was one of the first and most prominent authors who addressed the operation of these devices and analyzed the necessary requirements for generating work. He proposed a ``ratchet and pawl" device, presented as a microscopic-sized thermal ratchet model to illustrate the meaning and essence of the Second Law of Thermodynamics at the microscopic level.  He showed that when a system is in a single thermal bath the  ratchet and pawl is not capable of generating useful work or net movement. In other words, Feynman showed that in thermal equilibrium no net motion or work is achieved despite the anisotropy of the ratchet. However, by considering a temperature gradient, in combination with Brownian motion, it is indeed possible to induce a directed ratchet motion that can be transformed into work.

The model proposed by Feynman has been used as an inspiration in the study of a significant number of theoretical and experimental works on molecular motors and other micro-engine devices \cite{Magnasco1993,Reimann1996, julicher1997,  qian1997, Bier1997, Reimann2002, Hanggi2005, Lacoste2007,Perez2010, Goychuk2014, Tu2018, Hwang2019, Caballero2020, Gulyaev2020}. The so-called Brownian Motors \cite{Reimann1996, Reimann2002, Hanggi2005} are devices that are capable of rectifying fluctuations to produce useful work. They have a ratchet-like design where a spatial anisotropy, usually modeled as an asymmetric potential, is involved and, together with additional ingredients, takes the system out of equilibrium. Moreover, recent studies have been carried out exploring several aspects of these systems, to name a few: on the directed motion of cells in the total absence of gradients \cite{Caballero2020},  high-temperature ratchets driven by deterministic and stochastic fluctuations \cite{Rozenbaum2019}, flashing subdiffusive ratchets in viscoelastic media \cite{Kharchenko2012}, subdiffusive rocking ratchets in viscoelastic media \cite{Kharchenko2013}, fractional Brownian motors and stochastic resonance \cite{Goychuk2012}.  

Inspired by  Feynman's analysis, in 1993 Magnasco \cite{Magnasco1993} proposed a model called Forced Thermal Ratchet, capable of transporting a net flux of particles.  The model relies on an overdamped Brownian motion of a particle in an asymmetric periodic potential and subjected to the action of an external driving force. The author showed that the force and the broken symmetry are sufficient ingredients for particle transport. In particular, for a constant external driving, an analytical expression for stationary current is explicitly calculated, and in the case of square wave external modulation, the average rate of flow is obtained and plotted as a function of both the  external amplitude of the force and temperature. In the case of other time-dependent external driving the average rate of flows are calculated numerically.

Our present contribution focuses on the study of the forced thermal ratchet on an asymmetric periodic potential, in the high friction limit, which is modelled by a  Generalized Langevin Equation (GLE) that takes into account the cooperative effects of two  non-Markovian processes, namely, the friction memory kernel satisfying an Ornstein-Uhlenbeck-type process and the time-dependent driving force, also obeying an Ornstein-Uhlenbeck-like process.
 It is shown that in the overdamped regime, the presence of the memory heat bath induces a natural coupling between the internal noise correlation time with both, the first-order derivative at the position of the conservative force and first-order time derivative of the external force. This fact allows us to redefine the asymmetric potential proposed by Magnasco \cite{Magnasco1993}, for a continuous one but derivable in a small interval. In this case, our asymmetric and periodic potential is parabolic  just in a small interval characterized by parameter $\epsilon$. However, we show that in the limit of $\epsilon \to 0$, our theoretical result for the non-Markovian probability current is similar to the one reported in \cite{Magnasco1993}, except that the noise intensity as well as the external amplitude force are rescaled due to the presence of the noise correlation time. 
The exact analytical expression for such a non-Markovian probability current can be achieved by assuming a slow variation of the external force compared with the noise correlation and the inverse of the frequency driving.  For a square wave external modulation, we also obtain the behavior of the average rate of flow as a function of both the external amplitude force and bath temperature. We obtain novel results regarding the behavior of the non-Markovian stationary current when compared with the ones reported in the Markovian thermal ratchet  \cite{Magnasco1993}. As far as we are aware of, an exact explicit solution of the non-Markovian thermal ratchet with a slow variation of the amplitude driving force in an asymmetric periodic linear potential, has not been reported elsewhere. This is the main contribution in our work. In the case of either a finite time variation of the driving force \cite{Magnasco1993}, or for non-linear potential models, the stationary probability current can only be estimated numerically, as done, for instance, in \cite{Ibarra1997, Czernik1997, Bartussek1994}, but this is further the scope of the present paper.

\begin{figure}[hbtp]
        \centering
        \includegraphics[scale=0.9]{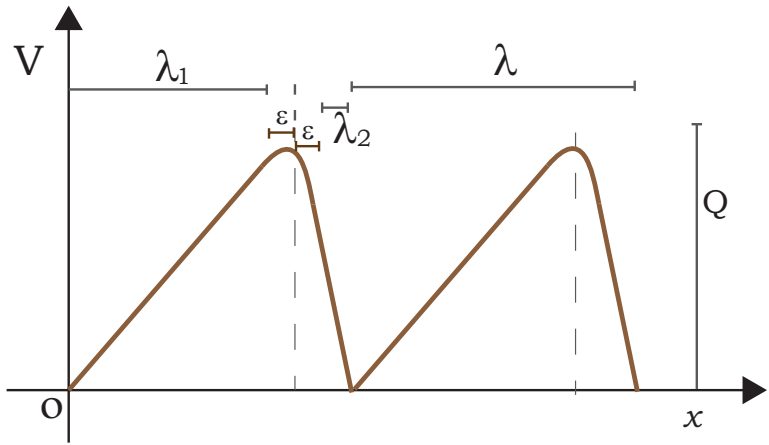}
        \caption {Piecewise linear potential $V(x)$ with a period $\lambda = \lambda_1 + \lambda_2$, and symmetry breaking amplitude $\Delta = \lambda_1 - \lambda_2$.}
        \label{potential}
    \end{figure}

This work is organized as follows: in Section \ref{sec2} we introduce the the non-Markovian forced thermal ratchet for an
arbitrary potential and then the theory is applied to  an specific potential as given in Fig. \ref{potential}. This potential is defined in three intervals and 
an explicit expression for the stationary probability current as function of the $\epsilon$ parameter can be calculated  

We then proceed, in the same Section, to compare our results with those of the Markovian case, and highlight in which scenarios a friction memory kernel may positively impact the transport properties of the thermal ratchet. We finalise this work with some conclusions and future prospects that can be read in Section \ref{sec3}.

\begin{eqnarray}
    V(x)= \begin{cases}
           {Q\over \lambda_1} x   &\text{ $x\,\in\, (0,\lambda_1 - \epsilon).$ }\\
           -{Q\over \lambda_1 \epsilon} x^2 + {2Q\over \epsilon}x - {Q\lambda_1 \over \epsilon} +Q &\text{ $x\,\in\, (\lambda_1 - \epsilon,\lambda_1 +\epsilon)$} \\
           -{Q\over \lambda_2} (x-\lambda_1) +Q   &\text{$x\,\in\, (\lambda_1+\epsilon,\lambda).$ }
           \end{cases}            \label{NMU}
\end{eqnarray}

\section{Non-Markovian forced thermal ratchet}
\label{sec2}

Consider a forced thermal ratchet model where the thermal interaction between the Brownian particle and the heat bath is finite-time correlated. In this case, the particle dynamics is described by a GLE with a friction memory kernel, which can be written as 
\be
m\dot v=-\int_0^t \gamma(t-t^{\prime}) \, v(t^{\prime}) dt^{\prime} +f(x) + F(t)+\mu(t) ,  \label{gle1}
\ee
where $\gamma(t)$ is the friction memory kernel, $f(x)$ is the force derived from a potential, 
$F(t)$ is a time-dependent driving force, and $\mu(t)$ is the fluctuating force with the following statistical properties 
\be
\langle \mu(t)\mu(t^{\prime})\rangle= k_{_B}T \gamma(t-t^{\prime}).  \label{corr}
\ee
In this work we assume that the memory kernel  satisfies an Ornstein-Uhlenbeck process with $\gamma(t-t^{\prime})={\gamma_0\over \tau} e^{-|t-t^{\prime}|/\tau}$, where $\gamma_0$ is the friction coefficient and $\tau$ the noise correlation time that quantifies the friction's memory. Hence, the GLE reads 
\be
m\dot v=-{\gamma_0\over \tau} \int_0^t  e^{-(t-t^{\prime})/\tau}\, v(t^{\prime}) dt^{\prime} +f(x)
+F(t)+\mu(t) . \label{gle2}
\ee
Next, to solve the problem we introduce the following change of variables: 
\ba
\eta(t)&=&-{\gamma_0\over \tau} \int_0^t  e^{-(t-t^{\prime})/\tau}\, v(t^{\prime}) dt^{\prime} +\mu(t) , \cr\cr
\mu(t)&=& {\sqrt{D}\over \tau} \int_0^t  e^{-(t-t^{\prime})/\tau}\, \xi(t^{\prime}) dt^{\prime} 
\label{eta} \ea
where $D=\gamma_0k_{_B}T$, and  the term $\xi(t)$ is a Gaussian white noise with zero mean value and correlation function  
$\langle \xi(t)\xi(t^{\prime})\rangle= 2 \delta(t-t^{\prime})$. In this case, 
the GLE transforms into the following coupled set of Langevin equations  
\ba
m\dot v&=&f(x)+F(t)+ \eta ,  \label{dotv} \\ 
\dot\eta&=&-{1\over\tau}\eta-{\gamma_0\over\tau}v+{\sqrt{D}\over\tau}\,\xi(t) . 
 \label{doteta}  \ea
In the overdamped regime, and taking $\gamma_0=1$ for simplicity, both equations can be combined into the following expression
\be
(1-f^{\prime}(x)\tau)\,\dot{x} = f(x) + F(t) +\tau{\dot F(t)} + \sqrt{D} \, \xi(t) .\label{dotx1} 
\ee
As it can be seen, in this regime the non-Markovian effect is coupled in a natural way to the rate of change of the driving force, as well as to the potential position derivative. To proceed further, we take the driving force rate to obey an Ornstein-Uhlenbeck-like process obeying the following equation  
\be
\dot{F} = \pm\,\omega F +  \sqrt{\sigma} \, \zeta(t),  \label{dotF} \ee
where $\zeta(t)$ is also a Gaussian white noise with zero mean value and correlation function $\langle \zeta(t)\zeta(t^{\prime})\rangle=2\delta(t-t^{\prime})$, $\sigma$ the noise intensity associated to the external force, and $\omega$ the inverse of the characteristic time of $F(t)$. In this case, Eq. (\ref{dotx1}) now reads
\be
(1-f^{\prime}(x)\tau)\,\dot{x} = f(x) + (1 \pm \omega \tau)F + \sqrt{D}\,\xi(t) +  \tau \sqrt{\sigma} \, \zeta(t).   \label{dotx2}
\ee
The Fokker-Planck equation (FPE) associated with this  effective Langevin equation can be written as 
\be
 {\partial \mathcal{P}(x,t)\over\partial t}= -{\partial \over \partial x} \bigg({F_e + f(x)\over 1-f^{\prime}(x)\tau} \,\mathcal{P}(x,t)
-{D_e\over 1-f^{\prime}(x)\tau }{\partial \mathcal{P}(x,t)\over\partial x} \bigg)  =-{\partial J(x,t)\over\partial x},   \label{fpe1}
\ee
where $J(x,t)$ is the probability current given  by 
\be
J(x,t) = {F_e + f(x)\over 1-f^{\prime}(x)\tau}\,
 \mathcal{P}(x,t)-{D_e\over 1-f^{\prime}(x)\tau}\,\frac{\partial \mathcal{P}(x,t)}{\partial x}. \label{Jo}
\ee
being $F_e(t)=(1{\pm}\omega\tau)F(t)$ and $D_e=k_{_B}T(1+\tau^2\,{\sigma\over k_BT})$, the corresponding effective amplitude force and effective noise intensity respectively. It is easy to verify that in the Markovian limit $\tau=0$, the probability current (\ref{Jo}) yields back to the the same result reported in \cite{Magnasco1993}. \\ 
For a stationary probability current the external amplitude force $F$ can be considered as a constant and in turn the effective amplitude force $F_e$, in such a way that 
\be
\frac{d \mathcal{P}(x)}{dx} - 
{F_e + f(x)\over D_e}\,
 \mathcal{P}(x)=- (1-f^{\prime}(x)\tau){\mathcal J}. \label{Joa}
\ee
where ${\mathcal J}= {J\over D_e}$. \\
We now can calculate the probability current for the specific piecewise potential as given in Fig. \ref{potential}. It clear that in the interval $(0,\lambda_1-\epsilon]$, the conservative force  $f(x)=-V^{\prime}(x)=-{Q\over\lambda_1}$ and $f^{\prime}(x)=0$, and therefore the probability current becomes 
\be
\frac{d \mathcal{P}_1(x)}{dx} - 
{1\over D_e}\bigg(F_e -{Q\over\lambda_1}\bigg)\,
 \mathcal{P}_1(x)=-{\mathcal J} . \label{Jo1}
\ee
For the interval $[\lambda_1-\epsilon, \lambda_1+\epsilon]$, the conservative force now reads $f(x)={2Q\over \lambda_1 \epsilon} x - {2Q\over \epsilon}$ and $f^{\prime}(x)= {2Q\over \lambda_1 \epsilon}$, so that the equation for the probability current becomes   
\be
\frac{d \mathcal{P}_2(x)}{dx} - 
{1\over D_e}\bigg(F_e +{2Q\over\lambda_1\epsilon}x -{2Q\over\epsilon} \bigg)\,
 \mathcal{P}_2(x)=-\bigg(1 -{2Q\over\lambda_1\epsilon} \tau\bigg) \mathcal J  . \label{Jo2}
\ee
And for the interval $[\lambda_1+\epsilon,\lambda)$, such that  that $f(x)=-V^{\prime}(x)={Q\over\lambda_2}$ and also  $f^{\prime}(x)=0$, so that the probability current reads 
\be
\frac{d \mathcal{P}_3(x)}{dx} - 
{1\over D_e}\bigg(F_e +{Q\over\lambda_1}\bigg)\,
 \mathcal{P}_3(x)=-{\mathcal J}. \label{Jo3}
\ee
From these equations it is straightforward to calculate  $\mathcal{P}_1(x)$, $\mathcal{P}_2(x)$, and $\mathcal{P}_3(x)$ for each interval.  

Moreover, due to the propagation of the probability density from the left to the right edges of the periodic piecewise potential, together with the normalization condition of the probability density, and taking the limit of $\epsilon \to 0$ (for a sawtooth-type potential) we can obtain 
the following analytical expression for the stationary non-Markovian probability current (see Appendix A)  
\ba
J&=& \frac{P^2_2 \sinh\mathcal X}{D_e\left({\lambda\over Q}\right)^2 [\cosh\mathcal Y-\cosh\mathcal X] 
-\left({\lambda\over Q}\right) P_1P_2\sinh\mathcal X} ,
 \label{J}   \ea
where 
\ba
{\mathcal X} &=& {\lambda F_e\over 2D_e}, \label{X} \qquad \qquad \qquad\qquad~~~~~
{\mathcal Y}={1\over D_e}\left(Q-{\Delta F_e\over2}\right) , \label{XY} \\
P_1&=& \Delta+{\lambda^2-\Delta^2\over 4}{F_e\over Q}, 
 \qquad \qquad 
 P_2=\left[ 1-{\Delta F_e\over 2Q}\right]^2- \left[{\lambda F_e\over 2Q}\right]^2. 
\label{p12} \ea
Here, the parameters $\Delta$, $\lambda$, and $Q$ characterise the shape of the piecewise linear potential, as shown in Fig. \ref{potential}. If the amplitude force 
is assumed to be a slowly-varying time square-wave signal with an amplitude of modulation $A$ (the modulation time is much greater than the noise correlation time $\tau$ and the inverse of the frequency driving $1/\omega$), then the effective force $F_e$ must also vary slowly. In this case average current reads 
\be
J_{Av}={1\over 2} [J(A)+J(-A) ]. \label{Jav} 
\ee

\subsection{Average current for a squared modulated signal}  
\begin{figure}
    \centering
    \includegraphics[width=8cm,height=6cm]{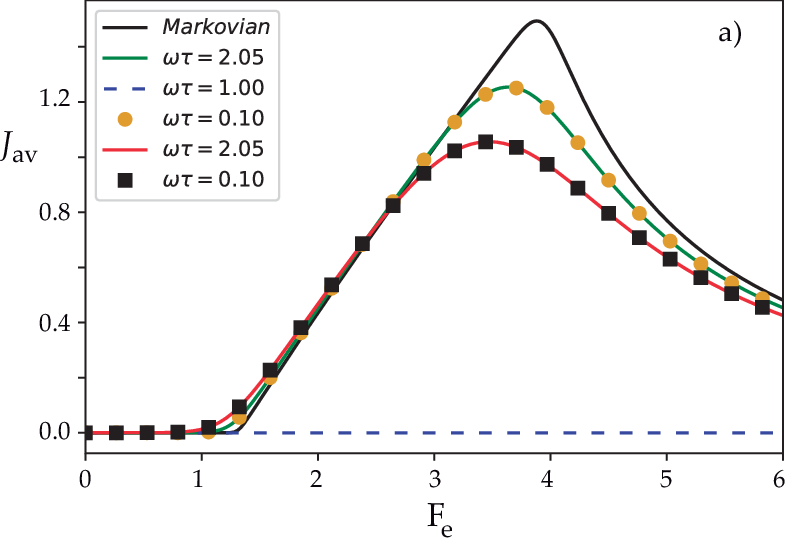}\quad\includegraphics[width=8cm,height=6cm]{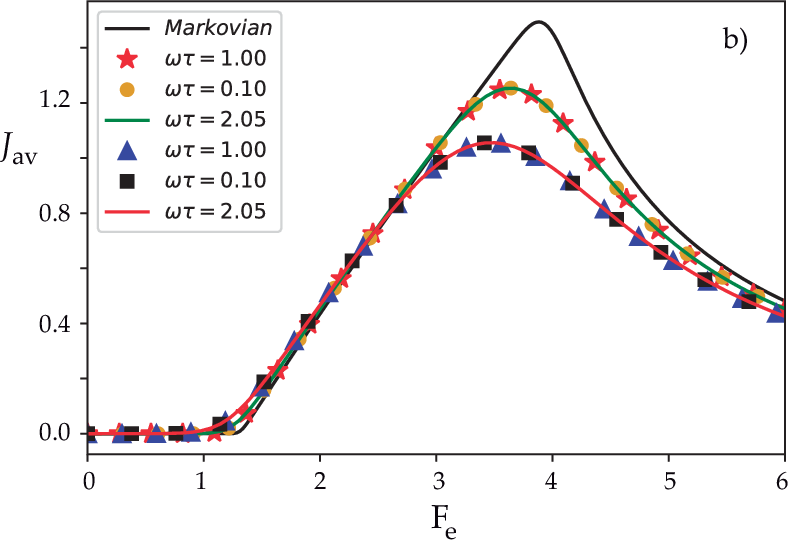}
    \caption{ $J_{Av}$ vs. $F_e$. a) $F_e=(1-\omega\tau)A$, $\tau=0.5$ (green line, golden bullets and dashed blue line) and $\tau=0.7$ (red line and black squares). b)  $F_e=(1+\omega\tau)F$,  $\tau=0.5$ (red stars, golden bullets and green line) and $\tau=0.7$ (blue triangles, black squares and the red line). In both cases $k_{_B}T=0.01$ and $\sigma=0.1$.}
    \label{Jav1}
\end{figure}
Let us proceed to analyze our main result, Eq. \eqref{J}, by comparing it with that corresponding to the Markovian case. The left panel in Fig. \ref{Jav1} shows the behaviour of the current $J_{Av}$ as a function of the absolute value of $F_e=(1-\omega\tau)A$ for three distinct cases corresponding to $\omega\tau>1$,  $\omega\tau<1$, and to $\omega\tau=1$. Moreover, for the first two cases we also show  the current for two values of $\tau$ equal to $0.5$ and $0.7$. As we can appreciate,  at a fixed value of $\tau$, the curves of the current collapse into the same universal curve, since these are plot against the effective force $F_e$. As one increases the value of $\tau$, the effective temperature increases, making the current overall weaker, even compared to the Markovian case. Remarkably however, the non-Markovian current activates sooner than the Markovian one, indicating that in e.g. a viscoelastic fluid one expects to see an initial enhancement of the current due to the positive correlations introduced by the memory kernel. Interestingly, the case $\omega\tau=1$ corresponds to a physical situation for which the average current is exactly null, meaning that the net current can be stopped by fine-tuning the value of $\omega$, without having to change the shape of the potential.\\
\begin{figure}
    \centering
    \includegraphics[scale=0.55]{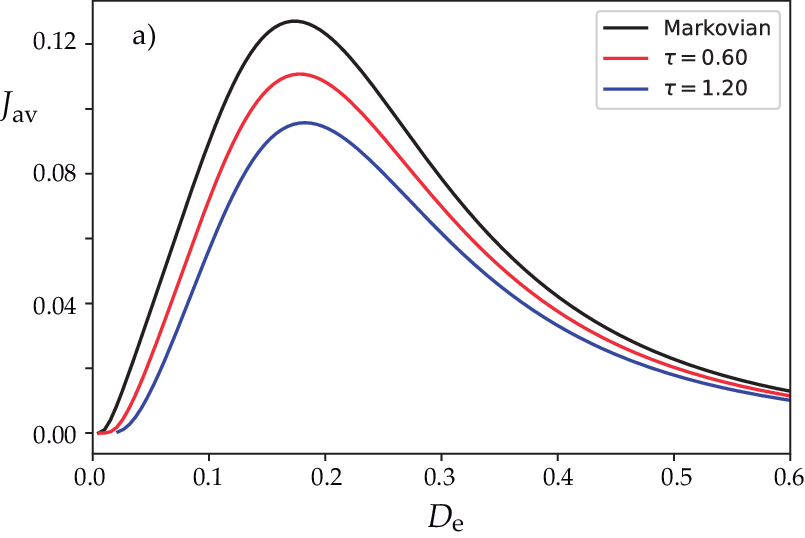}\includegraphics[scale=0.55]{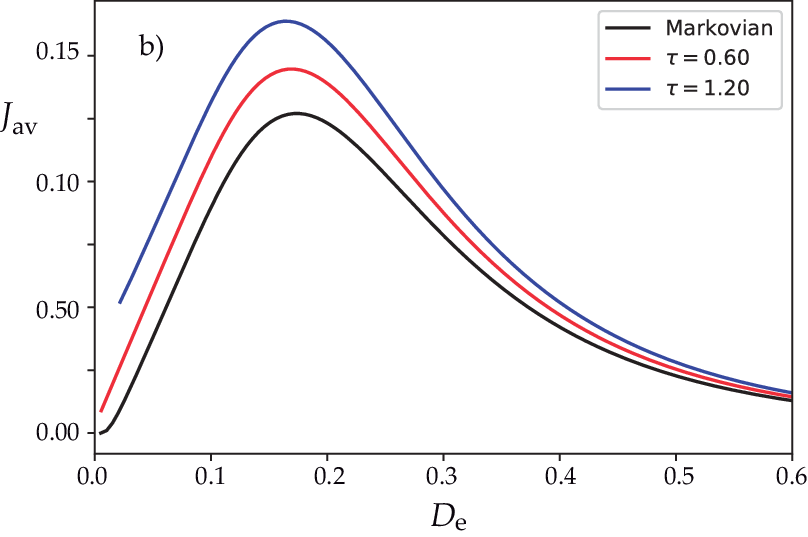}\par
    \caption{ $J_{Av}$ vs. $D_e$, for some values of $\tau$. a)  $F_e=(1-\omega \tau)A$ and b) $F_e=(1+\omega \tau)A$. In both cases  $A=1.0$,~ $\omega=0.10$, and  $\sigma=0.015$.}
    \label{Jav3}
\end{figure}
If we were to choose the positive sign in the Ornstein-Uhlenbeck process for the driving force, the effective amplitude force becomes $F_e=(1+\omega\tau)$. This means, as shown on the right panel of Fig. \ref{Jav1}, that for a fixed value of $\tau$, the non-Markovian current is universal when plotted as a function of $F_e$ and that by increasing $\tau$, as the effective temperature increases, the overall current becomes weaker, even to the corresponding Markovian case. In this case, however, there is no physical mechanism that produces a net zero current.

Similarly, we can plot the stationary current as a function of the effective noise $D_e=k_{_B}T(1+\tau^2 {\sigma\over k_{_B}T})$ for a fixed value of $\omega$, as reported in Fig. \ref{Jav3}. When the effective force $F_e=(1-\omega\tau)A$ the non-Markovian currents are smaller than the Markovian one. However, for $F_e=(1+\omega\tau)A$  there is an enhancement of the non-Markovian current with respect to the Markovian case.

\begin{figure}
    \centering
    \includegraphics[height=6cm,width=7.9cm]{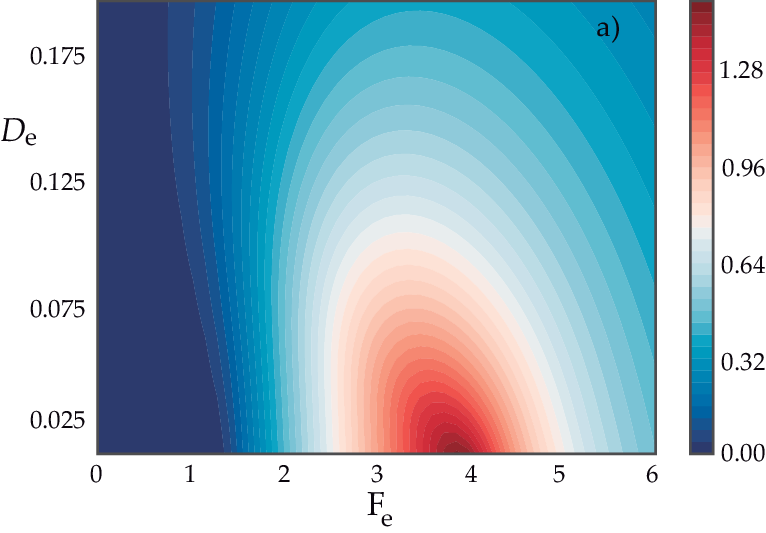}\includegraphics[height=6cm,width=7.9cm]{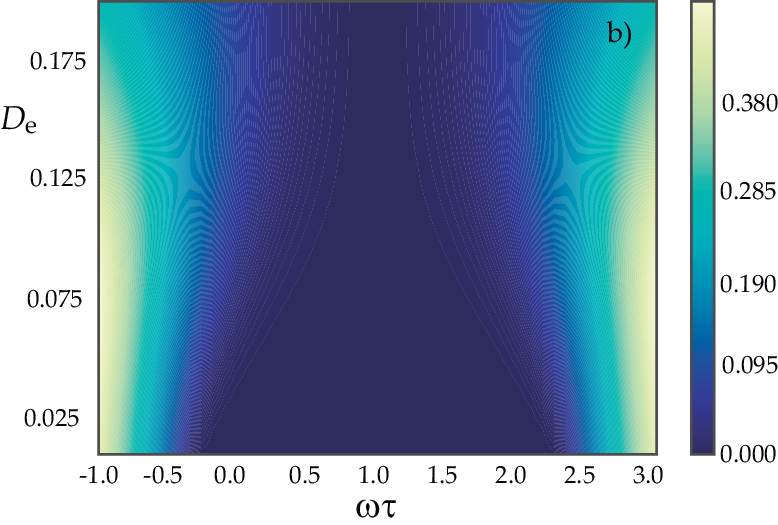}
    \caption{$J_{Av}$ vs. a) $(F_e,D_e)$  and b) $(\omega\tau,D_e)$. For a) $\omega\tau=2.05$, $\tau=0.5$ and $\sigma=0.015$, and b) $F_e=(1- \omega\tau)A$, with $A=1.0$,  $\tau=0.1$ and $\sigma=0.015$.}
    \label{Density_A}
\end{figure}

Finally, to visualise better the shape of the stationary non-Markovian current in terms of all its control parameters, the left and right panels in Fig. \ref{Density_A} show two density plots of the stationary current in the planes $(F_e,D_e)$ and $(\omega\tau,D_e)$, respectively. Notice from the density plot on the right panel that one may, fairly exquisitely, either switch on or off the current by tuning the characteristic time of the driving force either to a value $\omega\tau=1$ or away from it, respectively.

\section{Conclusions} 
\label{sec3}
In the present work, we have obtained in the high friction limit, an explicit expression of the probability current associated with a non-Markovian thermal ratchet characterized on a linear sawtooth-type potential (when the parameter $\epsilon\to 0$). This can be achieved if the Brownian particle in a thermal bath is described by a GLE with an OU friction memory kernel, and the time-dependent driving force also satisfies an OU-like process with a slow variation of the amplitude force. Our theoretical result (\ref{J}) is similar as the one reported in the  Markovian forced thermal ratchet \cite{Magnasco1993}, when the noise intensity $k_{_B}T$ and the amplitude force $F$ are  rescaled by  factors 
$1+ \tau^2 {\sigma\over k_{_B}T}$ and $1\pm\omega\tau$, respectively.

When comparing the non-Markovian average current to its Markovian counterpart, we notice that in some cases there is either an overall enhancement of the current or the current is activated before, indicating that the presence of a friction memory may be favorable in producing net forces in more realistic scenarios. Therefore, we believe that our findings open other possibilities of research, both theoretically and experimentally as, for instance, in the study of the effect that viscoelastic fluids may have in activated processes, similar to those recently reported in \cite{Kharchenko2012, Goychuk2012, Kharchenko2013,Goychuk2014}.

\begin{acknowledgments}
O. Contreras-Vergara thank the support from CONACyT-M\'exico. NSS thanks to SIP-IPN (M\'exico). The authors thank to M. D\'iaz-Segura for reading the manuscript.  
 \end{acknowledgments}

\appendix

\section{Probability current $J$}
The explicit solutions of $\mathcal{P}_1(x)$, $\mathcal{P}_2(x)$, and $\mathcal{P}_3(x)$, are given by
\ba
{\mathcal P}_1(x)&=&
{\mathcal{J} \over M} \bigg[ A e^{ax} +{M\over a} \bigg(1-e^{a x} \bigg) \bigg],  \label{P1-a}\\
{\mathcal  P}_2(x)&=&{\mathcal{J}  \over M} \bigg[ A e^{ B(x) +a(\lambda_1- \epsilon)} 
+{M\over a}\bigg( 1 - e^{a(\lambda_1-\epsilon)}  \bigg) e^{B(x)} 
   -\left(1 -{2Q\over\lambda_1\epsilon} \tau\right) \, M \,  K(x) \bigg] ,    \label{P2-a} \\
{\mathcal P}_3(x)&=&
{\mathcal{J}\over M} \bigg[A e^{a( \lambda_1- \epsilon) +b(x - \lambda_1-\epsilon) + B(\lambda_1+\epsilon) }  + {M\over a}\bigg( e^{ b(x - \lambda_1-\epsilon)} - 
e^{b(x - \lambda_1-\epsilon)+ a(\lambda_1- \epsilon) } \bigg) e^{ B(\lambda_1+\epsilon) } \cr\cr
&+& {M\over b} \bigg(  1 -  e^{b(x -\lambda_1-\epsilon)}  \bigg)  
- M {\left(1 -{2Q\over\lambda_1\epsilon} \tau\right)} \,  \, K(\lambda_1+\epsilon)   e^{b(x - \lambda_1-\epsilon)}  \bigg]   ,    \label{P3-a} 
 \ea
where $$a={1\over D_e}\left(F_e -{Q\over\lambda_1}\right), \quad b=  {1\over D_e} \left(F_e +{Q\over\lambda_2}\right)$$
\
$$K(x)= e^{a_1\left( x- { b_1\over 2 a_1}\right)^2} 
\int_{\lambda_1-\epsilon}^x e^{-a_1\left( y- { b_1\over 2 a_1}\right)^2} \, dy  ,
\qquad B(x)=a_1 x^2 + b_1 x  -c_1 -c_2.$$
\
\ba
a_1&=&{Q\over\ D_e \lambda_1\epsilon} , \qquad \qquad \qquad   b_1= {1\over D_e} 
\left(F_e - {2Q\over\epsilon} \right)  \cr\cr
c_1&=& {Q\over\ D_e \lambda_1\epsilon}(\lambda_1-\epsilon)^2,  \qquad 
c_2={1\over D_e} \left(F_e - {2Q\over\epsilon} \right)(\lambda_1-\epsilon). \nonumber
\ea
And
\ba
M&=&1- e^{\lambda \hat F_e -(a+b)\epsilon + B(\lambda_1+\epsilon) }    \label{M-a},\\ 
A&=& {1\over a}\bigg( e^{b(\lambda_2-\epsilon) + B(\lambda_1 + \epsilon ) }- 
e^{\lambda \hat F_e -(a+b)\epsilon + B(\lambda_1+\epsilon) } \bigg) \cr
&+& {1\over b}\bigg(1-e^{b(\lambda_2 -\epsilon)} \bigg)  
-\left(1 -{2Q\over\lambda_1\epsilon} \tau\right) \,  K(\lambda_1+\epsilon) e^{b(\lambda_2-\epsilon)} . \label{A-a}
\ea
with $\hat F_e=F_e/D_e$.
By applying the normalization condition for the total probability density ${\mathcal P}(x)$, it
can be shown that stationary probability current is given by ${\mathcal{J}}=M/(I_1+I_2+I_3)$, where
\ba
I_1&=& \int_0^{\lambda_1-\epsilon}
 \bigg[ A e^{ax} +{M\over a} \bigg(1-e^{a x} \bigg) \bigg]  dx,  \cr\cr
I_2&=&\int_{\lambda_1-\epsilon} ^{\lambda_1+\epsilon}  \bigg[ A e^{ B(x) +a(\lambda_1- \epsilon)} 
+{M\over a}\bigg( 1 - e^{a(\lambda_1-\epsilon)}  \bigg) e^{B(x)} 
 -\left(1 -{2Q\over\lambda_1\epsilon} \tau\right) \, M \,  K(x) \bigg]  dx,  \cr\cr
I_3&=&\int_{\lambda_1+\epsilon} ^{\lambda} \bigg[A e^{a( \lambda_1- \epsilon) +b(x - \lambda_1-\epsilon) + B(\lambda_1+\epsilon) }  + {M\over a}\bigg( e^{ b(x - \lambda_1-\epsilon)} - 
e^{b(x - \lambda_1-\epsilon)+ a(\lambda_1- \epsilon) } \bigg) e^{ B(\lambda_1+\epsilon) } \cr\cr
&+& {M\over b} \bigg(  1 -  e^{b(x -\lambda_1-\epsilon)}  \bigg)  
-  {\left(1 -{2Q\over\lambda_1\epsilon} \tau\right)} \, M  \, K(\lambda_1+\epsilon)   e^{b(x - \lambda_1-\epsilon)}  \bigg] dx  .    \label{I3} 
 \ea
\\
The expression $I_2$ can be separated in three terms
\ba
I_{21}&=& Ae^{a(\lambda_1 - \epsilon)}\int_{\lambda_1 -\epsilon}^{\lambda_1 + \epsilon}e^{B(x)} \,dx, \qquad I_{22} = {M\over a}\bigg( 1 - e^{a(\lambda_1-\epsilon)}  \bigg) \int_{\lambda_1 -\epsilon}^{\lambda_1 + \epsilon}e^{B(x)} \,dx \cr\cr
I_{23} &=& M\left(1 -{2Q\over\lambda_1\epsilon} \tau\right) \int_{\lambda_1 -\epsilon}^{\lambda_1 + \epsilon}K(x) \, dx.
\ea
According to the definitions given above we thus have 
\ba
B(\lambda_1+\epsilon)&=& a_1(\lambda_1+\epsilon)^2  + b(\lambda_1+\epsilon) -c_1-c_2  , \label{i1}  \\
\nonumber\\
 K(x)&=&{1\over 2}\sqrt{ \pi D_e \lambda_1\, \epsilon\over Q} e^{ Q\big[x-{(F_e -{2Q\over\epsilon})\lambda _1\epsilon \over 2 Q}\big]^2
\over D_e\lambda_1\epsilon} \bigg\{ {\rm Erf} \left[2Qx+(2Q-F_e\epsilon)\lambda_1\over 2\sqrt{Q D_e\lambda_1\epsilon}\right] \cr\cr
&-& {\rm Erf} \left[{-2Q\epsilon+ (4Q-F_e\epsilon)\lambda_1\over 2\sqrt{Q D_e \lambda_1\epsilon}} \right] \bigg\} \label{i2}  \ea
 We are interested to evaluate the stationary probability current in the limit of $\epsilon \to 0$. Due to the structure of the function $K(x)$, it is necessary to use the asymptotic behavior of the error function as given by
\begin{equation}
 \text{Erf}(y) = \frac{e^{-y^2}}{y\sqrt{\pi}}\left(  \sum_{n=0}^{\infty}\left(-1\right)\frac{1\cdot 3\cdot5\cdots (2n-1)}{(2y^2)^n}    \right),
\end{equation}
when $y\gg 1$. So, the first error function given in (\ref{i2}) can be written as
\begin{equation}
    {\rm Erf} \left[2Qx+(2Q-F_e\epsilon)\lambda_1\over 2\sqrt{Q D_e\lambda_1\epsilon}\right]=
    {\rm Erf}\left\{\sqrt{Q\over D_e\lambda_1}\left[ (x+\lambda_1){1\over\sqrt{\epsilon}}- {F_e\lambda_1\over 2Q}\, \sqrt{\epsilon} \right]\right\}.  \label{Erf1}
\end{equation}
In similar way the exponential in (\ref{i2}) can be written as 
\begin{equation}
  e^{ Q\big[x-{(F_e -{2Q\over\epsilon})\lambda _1\epsilon \over 2 Q}\big]^2\over D_e\lambda_1\epsilon} = e^{{Q\over D_e\lambda_1} \left[ (x+\lambda_1){1\over\sqrt{\epsilon}}- {F_e\lambda_1\over 2Q}\, \sqrt{\epsilon} \right]^2 }
\end{equation}
We now define the parameters 
$\alpha(x)=x+\lambda_1$, and $\beta={F_e \lambda_1\over 2Q}$, and therefore, as
 $\epsilon \to 0$ we thus have 
\begin{equation}
  e^{{Q\over D_e\lambda_1} \left[ \alpha(x){1\over\sqrt{\epsilon}}-\beta \sqrt{\epsilon} \right]^2 } {\rm Erf}\left\{\sqrt{Q\over D_e\lambda_1}\left [\alpha(x){1\over\sqrt{\epsilon}}- \beta \sqrt{\epsilon} \right]\right\} \approx{\sqrt{\epsilon}\over \sqrt{\pi}} \sqrt{D_e\lambda_1\over Q} {1\over\alpha(x)} . \label{Erf2} 
\end{equation}
For the second error function and as $\epsilon\to 0$ we get 
\ba
{\rm Erf} \left[{-2Q\epsilon+ (4Q-F_e\epsilon)\lambda_1\over 2\sqrt{Q D_e \lambda_1\epsilon}} \right]&=&
{\rm Erf}\left\{\sqrt{Q\over D_e\lambda_1}\left[ 2\lambda_1 {1\over\sqrt{\epsilon}}- \left(1+{F_e \lambda_1\over 2Q}\right)\, \sqrt{\epsilon} \right]\right\} \cr\cr
&\approx& {\sqrt{\epsilon}\over2 \sqrt{\pi} } \sqrt{D_e\over Q\lambda_1}e^{-{4 Q\lambda_1\over D_e}\, {1\over\epsilon} } ,  \label{Erf3}
\ea
which tends to zero as $\epsilon \to 0$. In this approximation $K(x)$ is given by 
\be
K(x)\approx {D_e \over 2Q}\,{\epsilon\over \alpha(x)}.
\ee
Therefore 
\begin{equation}
    \int_{\lambda_1 -\epsilon}^{\lambda_1 + \epsilon} K(x) \, dx \approx  {\epsilon \,D_e \over 2Q} \int_{\lambda_1 -\epsilon}^{\lambda_1 + \epsilon} \frac{dx}{x+\lambda_1} ={\epsilon \, D_e \over 2Q} \left[ \ln\left( { 2\lambda_1+\epsilon\over 2\lambda_1 -\epsilon }\right) \right]
\end{equation}
So that, $I_{23}$ reduces to 
\ba I_{23} &=& M\left(1 -{2Q\over\lambda_1\epsilon} \tau\right) \int_{\lambda_1 -\epsilon}^{\lambda_1 + \epsilon}K(x) \, dx \approx {D_e\over 2 Q}\, M \epsilon \left[ \ln\left( { 2\lambda_1+\epsilon\over 2\lambda_1 -\epsilon }\right) \right] \cr\cr
&-& {D_e \tau\over \lambda_1} M \left[ \ln\left( { 2\lambda_1+\epsilon\over 2\lambda_1 -\epsilon }\right) \right]   .  \label{I23} 
\ea
On the other side, it can be shown that 
\be
\int_{\lambda_1 -\epsilon}^{\lambda_1 + \epsilon}e^{B(x)} \,dx =\sqrt{ \pi D_e \lambda_1\, \epsilon\over 4Q} e^{-{\epsilon (F\lambda_1-2Q)^2\over 4Q D_e \lambda_1}} \left\{ {\rm Erfi} \left[(F\lambda_1+2Q)\sqrt{\epsilon}\over 2\sqrt{Q D_e\lambda_1}  \right] -
{\rm Erfi} \left[{(F\lambda_1-2Q)\sqrt{\epsilon}\over 2\sqrt{Q D_e\lambda_1} } \right] \right\}, \label{i3}  \ee
where ${\rm Erfi}(x)$ is the imaginary error function. From the results given above, it can be corroborated that  $\lim_{\epsilon\to 0} B(\lambda_1+\epsilon)=0$, ~$\lim_{\epsilon\to 0} K(\lambda_1+\epsilon)=0$, and according to Eqs. (\ref{I23}) and (\ref{i3}) 
\ba
&& \lim_{\epsilon\to 0} \int_{\lambda_1 -\epsilon}^{\lambda_1 + \epsilon}e^{B(x)} \,dx =0, \qquad \qquad \lim_{\epsilon\to 0}
I_{23}=0 .
\ea

Under these conditions, the non-Markovian average probability current for a square wave modulation is the same as the one given by Eq. (\ref{J}). Which corresponds 
to the sawtooth-type potential.

\bibliographystyle{unsrtnat}
\bibliography{biblioNM}

\begin{thebibliography}{22}
\providecommand{\natexlab}[1]{#1}
\providecommand{\url}[1]{\texttt{#1}}
\expandafter\ifx\csname urlstyle\endcsname\relax
  \providecommand{\doi}[1]{doi: #1}\else
  \providecommand{\doi}{doi: \begingroup \urlstyle{rm}\Url}\fi

\bibitem[Feynman et~al.((1966))Feynman, Leighton, and Sands]{Feynman1966}
R.~P. Feynman, R.~B. Leighton, and M.~Sands.
\newblock \emph{The Feynman Lectures on Physics}, volume~1.
\newblock (Addison-Wesley, Reading, MA), (1966).

\bibitem[Magnasco((1993))]{Magnasco1993}
M.~O. Magnasco.
\newblock Forced thermal ratchets.
\newblock \emph{Phys. Rev. Lett.}, {\bf{71}}:\penalty0 1477, (1993).
\newblock \doi{10.1103/PhysRevLett.71.1477}.

\bibitem[Reimann et~al.((1996))Reimann, Bartussek, H{\"a}ussler, and
  H{\"a}nggi]{Reimann1996}
P.~Reimann, R.~Bartussek, R.~H{\"a}ussler, and P.~H{\"a}nggi.
\newblock Brownian motors driven by temperature oscillations.
\newblock \emph{Phys. Lett. A}, {\bf 215}:\penalty0 26, (1996).
\newblock \doi{10.1016/0375-9601(96)00222-8}.

\bibitem[J{\"u}licher et~al.((1997))J{\"u}licher, Ajdari, and
  Prost]{julicher1997}
F.~J{\"u}licher, A.~Ajdari, and J.~Prost.
\newblock Modeling molecular motors.
\newblock \emph{Rev. Mod. Phys.}, {\bf 69}:\penalty0 1269, (1997).
\newblock \doi{10.1103/RevModPhys.69.1269}.

\bibitem[Qian((1997))]{qian1997}
H.~Qian.
\newblock A simple theory of motor protein kinetics and energetics.
\newblock \emph{Biophys. chem.}, {\bf 67}:\penalty0 263, (1997).
\newblock \doi{10.1016/S0301-4622(97)00051-3}.

\bibitem[Bier((1997))]{Bier1997}
M.~Bier.
\newblock Brownian ratchets in physics and biology.
\newblock \emph{Contemp. Phys.}, {\bf 38}:\penalty0 371, (1997).
\newblock \doi{10.1080/001075197182180}.

\bibitem[Reimann((2002))]{Reimann2002}
P.~Reimann.
\newblock Brownian motors: noisy transport far from equilibrium.
\newblock \emph{Phys. Rep.}, {\bf 361}:\penalty0 57, (2002).
\newblock \doi{10.1016/S0370-1573(01)00081-3}.

\bibitem[H{\"a}nggi et~al.((2005))H{\"a}nggi, Marchesoni, and Nori]{Hanggi2005}
P.~H{\"a}nggi, F.~Marchesoni, and F.~Nori.
\newblock Brownian motors.
\newblock \emph{Annalen der Physik}, {\bf 14}:\penalty0 51, (2005).
\newblock \doi{10.1002/andp.200410121}.

\bibitem[Lau et~al.((2007))Lau, Lacoste, and Mallick]{Lacoste2007}
A.~W.~C. Lau, D.~Lacoste, and K.~Mallick.
\newblock Nonequilibrium fluctuations and mechanochemical couplings of a
  molecular motor.
\newblock \emph{Phys. Rev. Lett.}, {\bf 99}:\penalty0 158102, (2007).
\newblock \doi{10.1103/PhysRevLett.99.158102}.

\bibitem[Perez-Carrasco and Sancho((2010))]{Perez2010}
R.~Perez-Carrasco and J.~M. Sancho.
\newblock Fokker-planck approach to molecular motors.
\newblock \emph{EPL}, {\bf 91}:\penalty0 60001, (2010).
\newblock \doi{10.1209/0295-5075/91/60001}.

\bibitem[Goychuk et~al.((2014))Goychuk, Kharchenko, and Metzler]{Goychuk2014}
I.~Goychuk, V.~Kharchenko, and R.~Metzler.
\newblock Molecular motors pulling cargos in the viscoelastic cytosol: how
  power strokes beat subdiffusion.
\newblock \emph{Chem. Phys.}, {\bf 16}:\penalty0 16524, (2014).
\newblock \doi{10.1039/C4CP01234H}.

\bibitem[Tu and Cao((2018))]{Tu2018}
Y.~Tu and Y.~Cao.
\newblock Design principles and optimal performance for molecular motors under
  realistic constraints.
\newblock \emph{Phys. Rev. E}, {\bf 97}:\penalty0 022403, (2018).
\newblock \doi{10.1103/PhysRevE.97.022403}.

\bibitem[Hwang and Karplus((2019))]{Hwang2019}
W.~Hwang and M.~Karplus.
\newblock Structural basis for power stroke vs. brownian ratchet mechanisms of
  motor proteins.
\newblock \emph{Proc. Natl. Acad. of Sci.}, {\bf 116}:\penalty0 19777, (2019).
\newblock \doi{10.1073/pnas.1818589116}.

\bibitem[Caballero et~al.((2020))Caballero, Kundu, and Reis]{Caballero2020}
D.~Caballero, S.~C. Kundu, and R.~L. Reis.
\newblock The biophysics of cell migration: biasing cell motion with feynman
  ratchets.
\newblock \emph{Biophys. J}, {\bf 1}, (2020).
\newblock \doi{10.35459/tbp.2020.000150}.

\bibitem[Gulyaev et~al.((2020))Gulyaev, Bugaev, Rozenbaum, and
  Trakhtenberg]{Gulyaev2020}
Y.~V. Gulyaev, A.~S. Bugaev, V.~M. Rozenbaum, and L.~I Trakhtenberg.
\newblock Nanotransport controlled by means of the ratchet effect.
\newblock \emph{Phys-Usp.}, {\bf 63}:\penalty0 311, (2020).
\newblock \doi{10.3367/UFNe.2019.05.038570}.

\bibitem[Rozenbaum et~al.((2019))Rozenbaum, Shapochkina, Teranishi, and
  Trakhtenberg]{Rozenbaum2019}
V.~M. Rozenbaum, I.~V. Shapochkina, Y.~Teranishi, and L.I. Trakhtenberg.
\newblock High-temperature ratchets driven by deterministic and stochastic
  fluctuations.
\newblock \emph{Phys. Rev. E}, {\bf 99}:\penalty0 012103, (2019).
\newblock \doi{10.1103/PhysRevE.99.012103}.

\bibitem[Kharchenko and Goychuk((2012))]{Kharchenko2012}
V.~Kharchenko and I.~Goychuk.
\newblock Flashing subdiffusive ratchets in viscoelastic media.
\newblock \emph{New J. Phys.}, {\bf 14}:\penalty0 043042, (2012).
\newblock \doi{10.1088/1367-2630/14/4/043042}.

\bibitem[Kharchenko and Goychuk((2013))]{Kharchenko2013}
V.~Kharchenko and I.~Goychuk.
\newblock Subdiffusive rocking ratchets in viscoelastic media: Transport
  optimization and thermodynamic efficiency in overdamped regime.
\newblock \emph{Phys. Rev. E}, {\bf 87}:\penalty0 052119, (2013).
\newblock \doi{10.1103/PhysRevE.87.052119}.

\bibitem[Goychuk and Kharchenko((2012))]{Goychuk2012}
I.~Goychuk and V.~Kharchenko.
\newblock Fractional brownian motors and stochastic resonance.
\newblock \emph{Phys. Rev. E}, {\bf 85}:\penalty0 051131, (2012).
\newblock \doi{10.1103/PhysRevE.85.051131}.

\bibitem[Ibarra-Bracamontes and Romero-Roch{\'\i}n((1997))]{Ibarra1997}
L.~Ibarra-Bracamontes and V.~Romero-Roch{\'\i}n.
\newblock Stochastic ratchets with colored thermal noise.
\newblock \emph{Phys. Rev. E}, {\bf 56}:\penalty0 4048, (1997).
\newblock \doi{10.1103/PhysRevE.56.4048}.

\bibitem[Czernik et~al.((1997))Czernik, Kula, {\L}uczka, and
  H{\"a}nggi]{Czernik1997}
T.~Czernik, J.~Kula, J.~{\L}uczka, and P.~H{\"a}nggi.
\newblock Thermal ratchets driven by poissonian white shot noise.
\newblock \emph{Phys. Rev. E}, {\bf 55}:\penalty0 4057, (1997).
\newblock \doi{10.1103/PhysRevE.55.4057}.

\bibitem[Bartussek et~al.((1994))Bartussek, H{\"a}nggi, and
  Kissner]{Bartussek1994}
R.~Bartussek, P.~H{\"a}nggi, and J.~G. Kissner.
\newblock Periodically rocked thermal ratchets.
\newblock \emph{EPL (Europhysics Letters)}, {\bf 28}:\penalty0 459, (1994).
\newblock \doi{10.1209/0295-5075/28/7/001}.

\end{thebibliography}

\end{document}